\documentclass[aps,pra,preprintnumbers,showpacs,tightenlines]{revtex4}

\usepackage{amssymb}
\usepackage{amsmath}
\usepackage{graphicx}
\usepackage{epsfig}
\usepackage{subfigure}
\usepackage{amsfonts}
\usepackage{CJK}

\begin{document}

\title{A proposal for implementing an $n$-qubit controlled-rotation gate
with three-level superconducting qubit systems in cavity QED}

\author{Chui-Ping Yang$^{1,2}$}

\address{$^1$Department of Physics, Hangzhou Normal University,
Hangzhou, Zhejiang 310036, China}

\address{$^2$State Key Laboratory of Precision Spectroscopy, Department of Physics,
East China Normal University, Shanghai 200062, China}

\begin{abstract}
We present a way for implementing an $n$-qubit controlled-rotation gate with
three-level superconducting qubit systems in cavity QED. The two logical states of a qubit are
represented by the two lowest levels of each system while a higher-energy
level is used for the gate implementation. The method operates essentially
by preparing a $W$ state conditioned on the states of the control qubits,
creating a single photon in the cavity mode, and then performing an
arbitrary rotation on the states of the target qubit with assistance of the
cavity photon. It is interesting to note that the basic operational steps
for implementing the proposed gate do not increase with the number $n$ of qubits,
and the gate operation time decreases as the number of qubits increases.
This proposal is quite general, which can be applied to various types
of superconducting devices in a cavity or coupled to a resonator.
\end{abstract}

 \maketitle
\date{\today}

\begin{center}
\textbf{I. INTRODUCTION}
\end{center}

Multiqubit controlled gates are of importance in constructing quantum
computation circuits and quantum information processing. A multiqubit
controlled gate can in principle be decomposed into the single-qubit and
two-qubit gates and thus can be built based on these elementary gates.
However, when using the conventional gate-decomposition protocols to
construct a multiqubit controlled gate~[1-3], the procedure usually becomes
complicated as the number of qubits increases. This is because single-qubit
and two-qubit gates, required for constructing a multiqubit controlled gate,
heavily depends on the number of qubits. Therefore, finding a more efficient
way to implement multiqubit controlled gates becomes important.

During the past few years, based on cavity QED technique, several
theoretical proposals for implementing an $n$-qubit controlled-phase gate
with ion traps, superconducting qubits coupled to a resonator or atoms
trapped in a cavity have been presented [4-10]. These previous works opened
a new way for the physical implementation of multiqubit controlled-phase (or
controlled-NOT) gates, which play a significant role in quantum information
processing, such as quantum algorithms [11,12] and quantum error-correction
protocols [13]. On the other hand, experimental realization of a three-qubit
controlled-phase gate in NMR quantum systems has been reported [14].
Moreover, a three-qubit quantum gate in trapped ions has been experimentally
demonstrated recently [15].

The existing proposals in [4-10] for realizing an $n$-qubit controlled-phase
gate can not be extended to realize an $n$-qubit controlled-$rotation$ gate
(denoted as controlled-$R$ gate below). Note that multiqubit controlled-$R$
gates are useful in quantum information processing. For instance, they can
be used to construct quantum circuits for: (i) general multiqubit gates
[16], (ii) preparation of arbitary pure quantum states
of multiple qubits [17], (iii) transformation of
quantum states of multiple qubits [18], and (iv) quantum error
correction [19]. In addition, multiqubit controlled-$R$ gates can be applied
to construct quantum circuits for implementation of quantum algorithms [20]
and quantum cloning [21], and so on. In this work, we will focus on how to
realize multiqubit controlled-$R$ gates with superconducting qubit systems.
As is well known, superconducting devices have appeared to be among the most
promising candidates for building quantum information processors recently,
due to their design flexibility, large-scale intergration, and compatibility
to conventional electronics.

We note that if an $n$-qubit controlled-$R$ gate is constructed by using the
conventional gate-decomposition protocols, $2^n-3$ two-qubit controlled
gates would be needed (for $n\geq 3$) [1]. Thus, assuming that realizing any
two-qubit controlled gate requires one-step operation only, at least $2^n-3$
steps of operations are required, which increase with the number $n$ of
qubits\textit{\ exponentially}. For instance, by using the conventional
gate-decomposing protocols, 29 basic operational steps are required to
implement a five-qubit controlled-$R$ gate, and 61 basic operational steps
are required to realize a six-qubit controlled-$R$ gate.

In the following, we will propose a way for implementing an $n$-qubit
controlled-$R$ gate with three-level superconducting qubit systems in cavity
QED. The method operates essentially based on this idea: prepare a $W$ state
conditional to the states of the control qubits, create a single photon in
the cavity mode, and then perform an arbitrary rotation on the states of the
target qubit with assistance of the cavity photon. The $W$ state used for
the gate implementation is defined as follows

\begin{equation}
\frac 1{\sqrt{n-1}}\sum P_z\left| 1\right\rangle ^{\otimes (n-2)}\left|
2\right\rangle,
\end{equation}
where $P_z$ is the symmetry permutation operator for qubit systems ($%
1,2,...,n-1$), $\sum P_z\left| 1\right\rangle ^{\otimes (n-2)}\left|
2\right\rangle $ denotes the totally symmetric state in which
one of qubit systems ($1,2,...,n-1$) is in the state $\left| 2\right\rangle $
while the remaining $n-2$ qubit systems are in the state $\left|
1\right\rangle .$ For instance, the $W$ state is $\frac 1{\sqrt{3}}\left(
\left| 112\right\rangle +\left| 121\right\rangle +\left| 211\right\rangle
\right) $ for $n-1=3.$ Note that $W$-class entangled states were originally
proposed by D\"ur W $et$ $al$. [22], which are useful in quantum information.

As shown below, our method only needs $7$ steps of operations, which is $%
independent$ of the number $n$ of qubits. Thus, when compared with the gate
operations required by the conventional gate-decomposition protocols, the
gate operations in this proposal are greatly simplified, especially when the
number $n$ of qubits is large. Furthermore, we note that the gate operation
time for this proposal decreases as the number of qubits increases. The
present proposal is quite general, which can be applied to various types of
superconducting devices in a cavity or coupled to a resonator.

This proposal requires adjustment of the level spacings of the qubit
systems. For solid-state qubit systems such as superconducting devices, the
level spacings can be rapidly adjusted (e.g., in 1 $\sim $ 2 nanosecond
timescale for superconducting qubits [23]), by varying the external
parameters (e.g., the external magnetic flux for superconducting charge
qubits, the flux bias or current bias in the case of superconducting phase
qubits and flux qubits, see e.g. [24-28]). It should be mentioned that
tuning the level spacings to have individual qubit systems coupled to or
decoupled from the cavity mode was earlier proposed for the physical
realization of quantum information processing with superconducting devices
(e.g., see [5,29-31]).

This paper is organized as follows. In Sec. II, we introduce the $n$-qubit
controlled-$R$ gate studied in this work. In Sec. III, we discuss how to
prepare the $W$ state conditioned on the states of the control qubits in
cavity QED. In Sec. IV, we present a way for implementing the $n$-qubit
controlled-$R$ gate with three-level superconducting qubit systems in a
cavity. In Sec. V, we give a brief discussion of the experimental
feasibility for implementing a six-qubit controlled-Hadamard gate with
superconducting devices coupled to a resonator. A concluding summary is
provided in Sec. VI.

\begin{center}
\textbf{II. }$N$\textbf{-QUBIT CONTROLLED-$R$ GATE}
\end{center}

\begin{figure}[tbp]
\begin{center}
\includegraphics[bb=203 394 348 592, width=5.0 cm, clip]{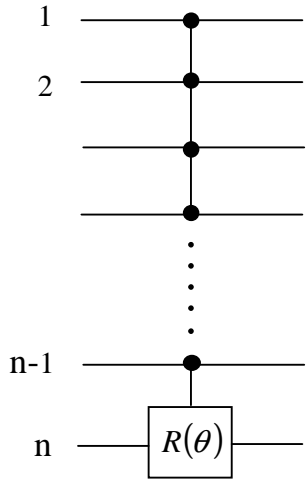} %
\vspace*{-0.08in}
\par
\end{center}
\caption{Schematic circuit of an $n$-qubit controlled-$R$ gate. If and only
if the $n-1 $ control qubits on the filled circles (qubits $1,2,....,$ and $%
n-1$) are all in the state $\left| 1\right\rangle $, a unitary rotation $%
R\left( \theta \right) $ is performed on the two logical states $\left|
0\right\rangle $ and $\left| 1\right\rangle $ of the target qubit (qubit $n$%
), by $\left| 0\right\rangle \rightarrow \cos \theta \left| 0\right\rangle
+\sin \theta \left| 1\right\rangle $ and $\left| 1\right\rangle \rightarrow
-\sin \theta \left| 0\right\rangle +\cos \theta \left| 1\right\rangle $.}
\label{fig:1}
\end{figure}

For $n$ qubits, there are a total number of $2^n$ computational basis
states, which form a set of complete orthogonal bases in a $2^n$-dimensional
Hilbert space of the $n$ qubits. A quantum controlled-$R$ gate of $n$ qubits
considered in this paper is defined by the following transformation:

\begin{equation}
\left| i_1i_2\cdot \cdot \cdot i_{n-1}\right\rangle \left| i_n\right\rangle
\rightarrow \left\{
\begin{array}{c}
\left| i_1i_2\cdot \cdot \cdot i_{n-1}\right\rangle R\left( \theta \right)
\left| i_n\right\rangle ,\;\text{if }\prod_{k=1}^{n-1}i_k=1 \\
\left| i_1i_2\cdot \cdot \cdot i_{n-1}\right\rangle \left| i_n\right\rangle
,\;\text{if }\prod_{k=1}^{n-1}i_k=0
\end{array}
\right.
\end{equation}
for all $i_1,i_2,\cdot \cdot \cdot ,i_n\in \left\{ 0,1\right\} .$ Here, the
subscripts $1,$ $2,...,$ and $n-1$ represent the $n-1$ control qubits ($%
1,2,...,n-1$) while the subscript $n$ represents the target qubit $n;$ and $%
\left| i_1i_2\cdot \cdot \cdot i_{n-1}\right\rangle \left| i_n\right\rangle $
is the $n$-qubit computational basis state. The operator $R\left( \theta
\right) $ is described by the following matrix
\begin{equation}
R\left( \theta \right) =\left(
\begin{array}{cc}
\cos \theta & -\sin \theta \\
\sin \theta & \cos \theta
\end{array}
\right)
\end{equation}
in a single-qubit computational subspace formed by the two logic states $%
\left| 0\right\rangle =\left( 1,0\right) ^T$and $\left| 1\right\rangle
=\left( 0,1\right) ^T$ of the target qubit $n$. It can be seen from Eq. (2)
that if and only if the $n-1$ control qubits ($1,2,...,n-1$) are all in the
state $\left| 1\right\rangle ,$ a unitary rotation $R\left( \theta \right) $
is performed on the two logic states $\left| 0\right\rangle $ and $\left|
1\right\rangle $ of the target qubit $n.$ The definition of the $n$-qubit
controlled-$R$ gate here is also shown in Fig.~1.

\begin{center}
\textbf{III. PREPARATION OF THE $W$ STATE CONDITIONED ON THE STATES OF THE
CONTROLS}
\end{center}

In this section, we will discuss how to prepare the $W$ state given in Eq.
(1), when the $n-1$ control qubits are initially in a computational basis
state $\left| 11...1\right\rangle .$ For the gate purpose, the remaining $%
2^{n-1}-1$ computational states of the $n-1$ control qubits need to be not
affected during the $W$ state preparation. In this section, we will also
give a discussion on how this can be achieved.

\begin{figure}[tbp]
\includegraphics[bb=144 234 415 485, width=8.0 cm, clip]{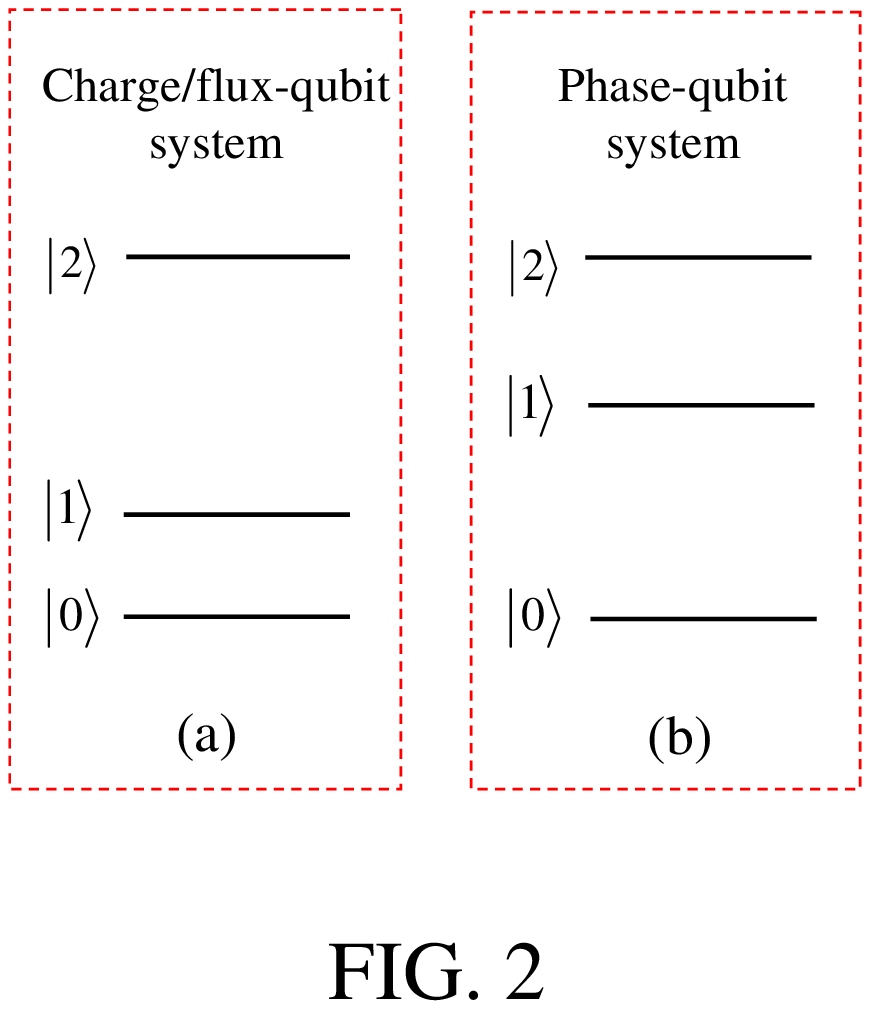} %
\vspace*{-0.08in}
\caption{(Color online) Illustration of three-level superconducting qubit
systems. In (a), the level spacing between the two upper levels is larger
than that between the two lower levels. In (b), vice versa.}
\label{fig:2}
\end{figure}

The superconducting qubit systems have the three levels shown in Fig.~2.
Note that the three-level structure in Fig.~2(a) applies to superconducting
charge-qubit or flux-qubit systems [24,25] and the one in Fig.~2(b) applies
to phase-qubit systems [26,27]. In addition, the three-level structure in
Fig.~2(b) is also available in atoms. In Fig.~2, the transition between the
two lowest levels $\left| 0\right\rangle $ and $\left| 1\right\rangle $ is
assumed to: (i) be forbidden due to the selection rules, (ii) very week due
to the potential barrier between the two lowest levels, or (iii) highly
detuned (decoupled) from the cavity mode during the gate operation.
Throughout this paper, the two logic states of a qubit are represented by
the two lowest levels $\left| 0\right\rangle $ and $\left| 1\right\rangle .$

To simplify our presentation, we will restrict our discussion to the
three-level structure in Fig.~2(a). However, it should be mentioned that the
results presented in this section and the method proposed in next section
for the gate implementation are applicable to the quantum systems with the
three-level structure depicted in Fig.~2(b).

\begin{figure}[tbp]
\begin{center}
\includegraphics[bb=232 416 385 574, width=7.0 cm, clip]{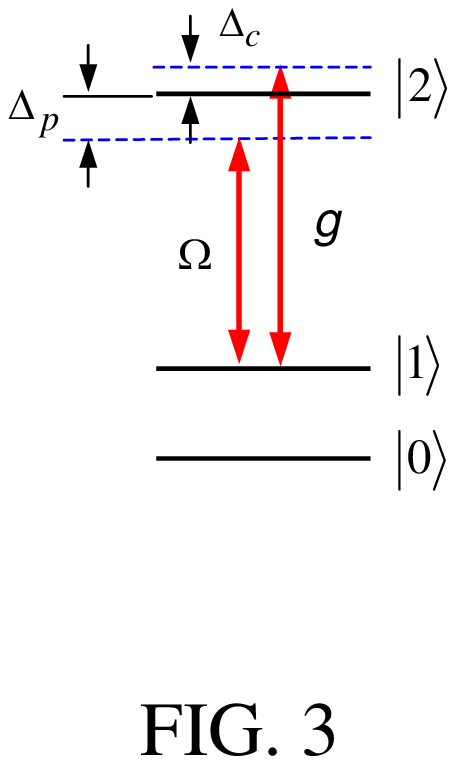} %
\vspace*{-0.08in}
\par
\end{center}
\caption{(Color online) The transition between the two lowest levels is
forbidden due to the selection rules, very weak due to the potential barrier
between the two lowest levels, or highly detuned (decoupled) from the cavity
mode. The $\left| 1\right\rangle \leftrightarrow \left| 2\right\rangle $
transition is non-resonantly coupled to the cavity mode with a detuning $%
\Delta _c$, and non-resonantly coupled to the classical pulse with a
detuning $\Delta_p$.}
\label{fig:3}
\end{figure}

Consider $n-1$ three-level superconducting qubit systems ($1,2,...,n-1$) in
a single-mode cavity. The $n-1$ systems play the role of controls in
realization of the $n$-qubit controlled-$R$ gate, discussed in next section.
Assume that the cavity mode is coupled to the $\left| 1\right\rangle
\leftrightarrow $ $\left| 2\right\rangle $ transition but does not affect
the level $\left| 0\right\rangle $ (Fig.~3), which can be achieved by prior
adjustment of the level spacings [24-28]. The Hamiltonian is given by
(assuming $\hbar =1$)
\begin{equation}
H=\omega _0S_z+\omega _ca^{+}a+g\left( a^{+}S^{-}+aS^{+}\right) ,
\end{equation}
where $S_z=\frac 12\sum_{j=1}^{n-1}(\left| 2\right\rangle _j\left\langle
2\right| -\left| 1\right\rangle _j\left\langle 1\right|
),S^{+}=\sum_{j=1}^{n-1}\left| 2\right\rangle _j\left\langle 1\right| ,$ $%
S^{-}=\sum_{j=1}^{n-1}\left| 1\right\rangle _j\left\langle 2\right| ,$ $a^{+}
$ and $a$ are the photon creation and annihilation operators for the cavity
mode, $\omega _c$ is the cavity-mode frequency, $\omega _0$ is the $\left|
1\right\rangle \leftrightarrow \left| 2\right\rangle $ transition frequency$,
$ and $g$ is the coupling constant between the cavity mode and the $\left|
1\right\rangle \leftrightarrow \left| 2\right\rangle $ transition.

In the case when the detunning $\Delta _c=\omega _c-\omega _0\gg g\sqrt{%
\overline{n}+1}$ with $\overline{n}$ being the mean photon number of the
cavity mode, the Hamiltonian (4) can be rewritten as follows [32]
\begin{equation}
H=\omega _0S_z+\omega _ca^{+}a-\lambda \left[ \sum_{j=1}^{n-1}(\left|
2\right\rangle _j\left\langle 2\right| -\left| 1\right\rangle _j\left\langle
1\right| )a^{+}a\right] -\lambda S^{+}S^{-},
\end{equation}
where $\lambda =g^2/\Delta _c.$ The third term describes the photon-number
dependent Stark shift while the last term describes the dipole coupling
among the $n-1$ qubit systems. If the cavity mode is initially in the vacuum
state $\left| 0\right\rangle $, the Hamiltonian (5) reduces to
\begin{equation}
H_0=\omega _0S_z-\lambda S^{+}S^{-}.
\end{equation}

If the $n-1$ control qubits are initially in the computational basis state $%
\left| 11...1\right\rangle $, i.e., the Dicke state $\left|
J,-J\right\rangle $ with $J=\left( n-1\right) /2,$ they evolve within the
symmetric Dicke subspace spanned by $\left\{ \left| J,-J\right\rangle
,\left| J,-J+1\right\rangle ,...,\left| J,J\right\rangle \right\} .$ Here,
the state $\left| J,-J+k\right\rangle $ with $k=0,1,...,n-1$ is a symmetric
Dicke state with $k$ systems being in the state $\left| 2\right\rangle $
while $n-k-1$ systems being in the state $\left| 1\right\rangle $. The Dicke
state $\left| J,-J+k\right\rangle $ is given by

\begin{equation}
\left| J,-J+k\right\rangle =\frac 1{\sqrt{n-1}}\sum P_z\left| 1\right\rangle
^{\otimes (n-k-1)}\left| 2\right\rangle ^{\otimes k},
\end{equation}
where $P_z$ is the symmetry permutation operator for systems ($1,2,...,n-1$%
), $\sum P_z\left| 1\right\rangle ^{\otimes (n-k-1)}\left| 2\right\rangle
^{\otimes k}$ denotes the totally symmetric state in which $n-k-1$ of
systems ($1,2,...,n-1$) are in the state $\left| 1\right\rangle $ while the
remaining $k$ systems are in the state $\left| 2\right\rangle .$

\begin{figure}[tbp]
\begin{center}
\includegraphics[bb=143 500 458 736, width=9.0 cm, clip]{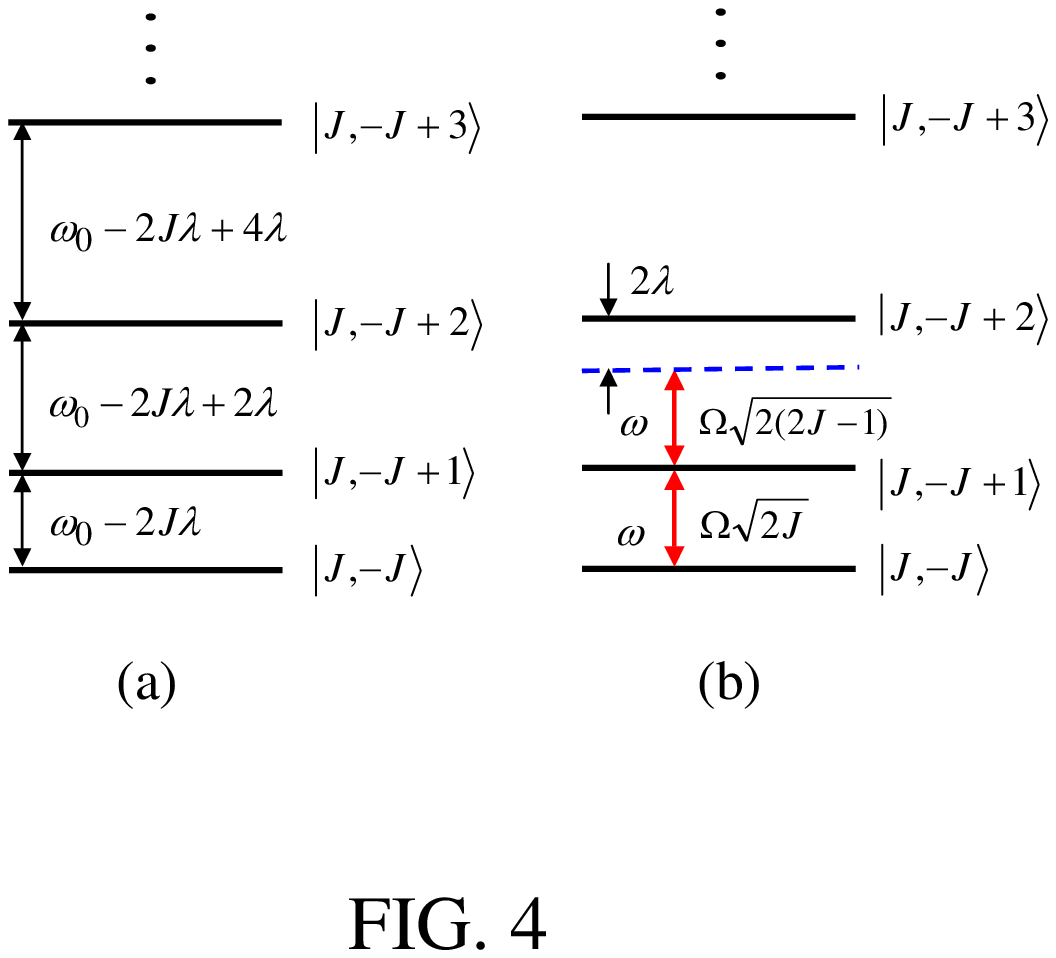} %
\vspace*{-0.08in}
\par
\end{center}
\caption{(Color online) (a) Non-identical level spacings for the energy
levels. The level spacings between two adjacent levels become wider by $%
2\lambda $ as the energy levels go up. (b) Illustration of the pluse (with
frequency $\omega=\omega_0-2J\lambda$) resonantly coupled to the transition
between the two lowest Dicke states $\left| J,-J\right\rangle $ and $\left|
J,-J+1\right\rangle $ but detuned from the $\left| J,-J+1\right\rangle
\leftrightarrow \left| J,-J+2\right\rangle $ transition with a detuning $%
2\lambda $. }
\label{fig:4}
\end{figure}

One can check that the Hamiltonian $H_0$ has the following properties

\begin{equation}
H_0\left| J,-J+k\right\rangle =\varepsilon _k\left| J,-J+k\right\rangle ,
\end{equation}
with
\begin{equation}
\varepsilon _k=\omega _0\left( -J+k\right) -k\left( 2J-k+1\right) \lambda .
\end{equation}
Eq.~(8) demonstrates that the Dicke state $\left| J,-J+k\right\rangle $ is
the eigenstate of the Hamiltonian $H_0$ with the eigenvalue $\varepsilon _k.$
The energy-level spacing between $\left| J,-J+k\right\rangle $ and $\left|
J,-J+k+1\right\rangle $ is $\varepsilon _{k+1}-\varepsilon _k=\omega
_0-2\left( J-k\right) \lambda $, depending on the excitation number of the
state $\left| J,-J+k\right\rangle $. It can be seen that the energy level
spacings in the symmetric Dicke subspace are unequal [Fig.~4(a)]. For the
detailed discussion, see Ref.~[33].

To prepare the $W$ state of Eq. (1), we now apply an external driving pulse
(with frequency $\omega $) to the $n-1$ systems ($1,2,...,n-1$). Suppose
that the pulse is coupled to the $\left| 1\right\rangle \leftrightarrow
\left| 2\right\rangle $ transition but far-off resonant with the transition
between any other two levels of each system (Fig.~3). Thus, the interaction
Hamiltonian between the pulse and the $n-1$ systems is given by
\begin{equation}
H_{sp}=\Omega \left( e^{-i\omega t}S^{+}+e^{i\omega t}S^{-}\right) ,
\end{equation}
where $\Omega $ is the Rabi frequency of the pulse. The Hamiltonian for the
whole system is
\begin{equation}
\widetilde{H}=H_0+H_{sp}
\end{equation}
$.$

Performing the transformation $U=e^{i\omega tS_z}$, we obtain the engineered
Hamiltonian in the symmetric Dicke subspace
\begin{eqnarray}
H^{^{\prime }} &=&U\widetilde{H}U^{+}-\omega S_z  \nonumber \\
&=&\sum_{k=0}^{n-1}\delta _k\left| J,-J+k\right\rangle \left\langle
J,-J+k\right|  \nonumber \\
&&+\sum_{k=0}^{n-2}\Omega _k\left( \left| J,-J+k+1\right\rangle \left\langle
J,-J+k\right| +\text{H.c.}\right) ,
\end{eqnarray}
where
\begin{eqnarray}
\Omega _k &=&\Omega \sqrt{\left( 2J-k\right) \left( k+1\right) },  \nonumber
\\
\delta _k &=&\omega _0\left( -J+k\right) -k\left( 2J-k+1\right) \lambda
-\omega \left( -J+k\right) .
\end{eqnarray}
Assume that the applied pulse is resonant with the transition between the
Dicke states $\left| J,-J\right\rangle $ and $\left| J,-J+1\right\rangle $
[Fig.~4(b)]$.$ Namely, the pulse frequency $\omega $ is set by $\omega
=\omega _0-2J\lambda ,$ i.e., $\Delta _p=\omega _0-\omega =2J\lambda $
(Fig.~3). Discarding the constant energy $-2J^2\lambda $ we have $\delta
_0=\delta _1=0$ and $\delta _k=k\left( k-1\right) \lambda $ $(k\geq 2)$.
Hence the detuning of the pulse frequency with the transition frequency
between the two energy levels $\left| J,-J+1\right\rangle $ and $\left|
J,-J+2\right\rangle $ is $2\lambda $ [Fig.~4(b)]. Therefore, if we have $%
\Omega \sqrt{2\left( 2J-1\right) }\ll 2\lambda ,$ which is guaranteed by
setting
\begin{equation}
\Omega \sqrt{n-1}\ll \lambda ,
\end{equation}
then the transition between the two Dicke states $\left| J,-J+1\right\rangle
$ and $\left| J,-J+2\right\rangle $ is negligible due to far-off resonance
with the pulse. As a result, when the systems ($1,2,...,n-1$) are initially
in the Dicke state $\left| J,-J\right\rangle $ or $\left|
J,-J+1\right\rangle $, the Dicke state $\left| J,-J+2\right\rangle $ will
not be excited by the pulse and therefore no transition from the state $%
\left| J,-J+2\right\rangle $ to any one of the Dicke states \{$\left|
J,-J+3\right\rangle ,\left| J,-J+4\right\rangle ,...,\left| J,J\right\rangle
$\} occurs. The Hamiltonian $H^{\prime }$ thus reduces to

\begin{equation}
H^{\prime }=\Omega \sqrt{2J}\left( \left| J,-J+1\right\rangle \left\langle
J,-J\right| +\text{H.c.}\right) .
\end{equation}

It is straightforward to show from Eq.~(15) that the states $\left|
J,-J\right\rangle $ and $\left| J,-J+1\right\rangle $ evolve as follows
\begin{equation}
\left| J,-J\right\rangle \rightarrow \cos (\sqrt{2J}\Omega t)\left|
J,-J\right\rangle -i\sin (\sqrt{2J}\Omega t)\left| J,-J+1\right\rangle ,
\end{equation}
\begin{equation}
\left| J,-J+1\right\rangle \rightarrow \cos (\sqrt{2J}\Omega t)\left|
J,-J+1\right\rangle -i\sin (\sqrt{2J}\Omega t)\left| J,-J\right\rangle .
\end{equation}
Based on Eq. (1) and Eq. (7), it can be seen that the Dicke state $\left|
J,-J+1\right\rangle $ here is the $W$ state defined in Eq. (1).

From Eq.~(16), it can be seen that when the control qubits ($1,2,...,n-1$)
are initially in the state $\left| J,-J\right\rangle $ (i.e., the
computational state $\left| 11..1\right\rangle $), the $W$ state $\left|
J,-J+1\right\rangle $ is prepared through a transformation $\left|
J,-J\right\rangle \rightarrow -i\left| J,-J+1\right\rangle $ after a pulse
duration $t=\pi /(2\sqrt{2J}\Omega )$.

For the gate implementation below, we will need to transform the prepared $W$
state $\left| J,-J+1\right\rangle $ back to the state $\left|
J,-J\right\rangle .$ Eq.~(17) shows that this can be achieved by applying
the same pulse to the systems ($1,2,...,n-1$) for a time $t=\pi /(2\sqrt{2J}%
\Omega )$, via the transformation $\left| J,-J+1\right\rangle \rightarrow
-i\left| J,-J\right\rangle $.

The remaining $\left( 2^{n-1}-1\right) $ computational basis states $\left|
i_1i_2...i_{n-1}\right\rangle $ of the $n-1$ control qubit systems can be
classified into a set of Dicke states $\left\{ \left|
J-l/2,-(J-l/2)\right\rangle \right\} $ with $l=1,2,...,n-1.$ For instance,
the $(n-1)$-qubit computational basis states $\left| 11...110\right\rangle $
and $\left| 11...100\right\rangle $ can be written in term of the Dicke
states $\left| J-1/2,-(J-1/2)\right\rangle $ (for $l=1$) and $\left|
J-1,-(J-1)\right\rangle $ (for $l=2$), respectively. To see how the set of
Dicke states $\left\{ \left| J-l/2,-(J-l/2)\right\rangle \right\} $ here not
to be affected during the $W$ state preparation, let us focus on a Dicke
state $\left| J-l/2,-(J-l/2)\right\rangle $ ($l\neq 0$), and discuss how to
make this Dicke state unaffected by the pulse.

The level spacing between the Dicke states $\left|
J-l/2,-(J-l/2)\right\rangle $ and $\left| J-l/2,-(J-l/2)+1\right\rangle $ is
given by
\begin{equation}
\widetilde{\epsilon }_1-\widetilde{\epsilon }_0=\omega _0-(2J-l)\lambda .
\end{equation}
Therefore, the detuning of the pulse frequency from the transition frequency
between the two Dicke states $\left| J-l/2,-(J-l/2)\right\rangle $ and $%
\left| J-l/2,-(J-l/2)+1\right\rangle $ would be $\omega -(\widetilde{%
\epsilon }_1-\widetilde{\epsilon }_0)=-l\lambda $. The pulse Rabi frequency
between the two Dicke states $\left| J-l/2,-(J-l/2)\right\rangle $ and $%
\left| J-l/2,-(J-l/2)+1\right\rangle $ is $\Omega \sqrt{2J-l}$, which can be
seen from the expression of $\Omega _k$ in Eq.~(13) (for the present case, $%
k=0$ and $J$ is replaced by $J-l/2$)$.$ If the large detuning $\Omega \sqrt{%
2J-l}\ll l\lambda $ is met, the transition between the two Dicke states $%
\left| J-l/2,-(J-l/2)\right\rangle $ and $\left|
J-l/2,-(J-l/2)+1\right\rangle $ can be neglected due to far-off resonance
with the pulse. Thus, the pulse does not excite the Dicke state $\left|
J-l/2,-(J-l/2)+1\right\rangle $ when the control qubits ($1,2,...,n-1$) are
initially in the state $\left| J-l/2,-(J-l/2)\right\rangle .$

Note that for any $l\in \left\{ 1,2,...,n-1\right\} ,$ we have $\Omega \sqrt{%
2J-l}\ll l\lambda $ when $\Omega \sqrt{n-1}\ll \lambda .$ Therefore, as long
as the condition in Eq.~(14) is satisfied, the Dicke state $\left|
J-l/2,-(J-l/2)\right\rangle $ with any given $l=1,2,...,$ or $n-1$ will not
be affected by the pulse, i.e., the rest $\left( 2^{n-1}-1\right) $
computational basis states of the control qubit systems ($1,2,...,n-1$)
remain unchanged during the pulse.

\begin{center}
\textbf{IV. IMPLEMENTATION OF AN }$N$\textbf{-QUBIT CONTROLLED-}$R$\textbf{\
GATE} \textbf{IN CAVITY QED}
\end{center}

Let us now consider $n$ superconducting qubit systems ($1,2,...,n$) placed
in a cavity or coupled to a resonator. Each system has the three-level
configuration. Initially, the transition between any two levels of each
system is highly detuned (decoupled) from the cavity mode [Fig. 5(a, a$%
^{\prime }$)], which can be achieved via prior adjustment of the level
spacings. In addition, assume that the cavity mode is initially in the
vacuum state $\left| 0\right\rangle _c.$

\begin{figure}[tbp]
\begin{center}
\includegraphics[bb=166 124 413 743, width=7.0 cm, clip]{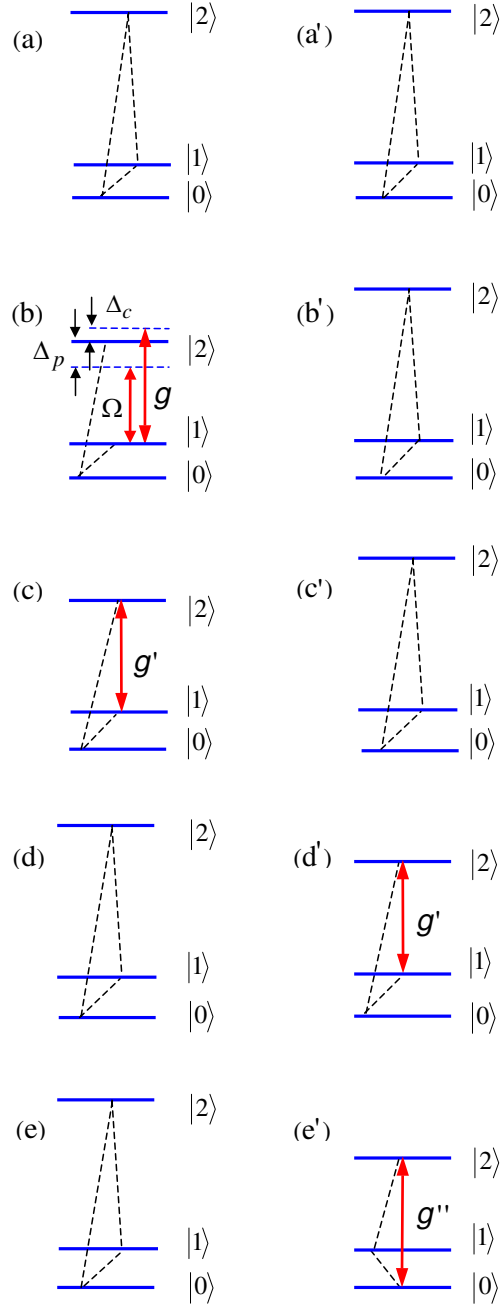} %
\vspace*{-0.08in}
\par
\end{center}
\caption{(Color online) The level structures of the systems ($1,2,...,n$)
during the gate preparation. Figures on the left side represent the level
structures for systems ($1,2,...n-1$), while figures on the right size
represent the level structures of system $n$. Here, $g$ is the
non-resonantly-coupling constant between the cavity mode and the $\left|
1\right\rangle \leftrightarrow \left| 2\right\rangle $ transition, $%
g^{\prime }$ is the resonantly-coupling constant between the cavity mode and
the $\left| 1\right\rangle \leftrightarrow \left| 2\right\rangle $
transition, and $g^{\prime \prime }$ is the resonantly-coupling constant
between the cavity mode and the $\left| 0\right\rangle \leftrightarrow
\left| 2\right\rangle $ transition. In addition, the transition between any
two levels linked by a dashed line is highly detuned (decoupled) from the
cavity mode and/or the pulse.}
\label{fig:5}
\end{figure}

The procedure for implementing the $n$-qubit controlled-$R$ gate is listed
as follows:

Step (i): Leave the level structure of system $n$ unchanged [Fig. 5(b$%
^{\prime }$)] while adjusting the level spacings of systems ($1,2,...,n-1$)
such that the $\left| 1\right\rangle \leftrightarrow \left| 2\right\rangle $
transition of each of systems ($1,2,...,n-1$) is non-resonantly coupled to
the cavity mode, with a detuning $\Delta _c$ [Fig.~5(b)]. Then, apply a
pulse to systems ($1,2,...,n-1$) for a duration $t_1=\pi /(2\sqrt{2J}\Omega
) $ [Fig.~5(b)]. As discussed in Sec. III, when the systems ($1,2,...,n-1$)
are initially in the computational basis state $\left| 11...1\right\rangle ,$
the $W$ state $\left| J,-J+1\right\rangle $ is created after the pulse, via
a transformation $\left| J,-J\right\rangle \rightarrow $ $-i\left|
J,-J+1\right\rangle $. Note that the cavity mode remains in the vacuum state
during the operation of this step, as shown in Sec. III.

Step (ii): Leave the level structure of system $n$ unchanged [Fig.~5(c$%
^{\prime }$)] while adjusting the level spacings of systems ($1,2,...,n-1)$
such that the $\left| 1\right\rangle \leftrightarrow \left| 2\right\rangle $
transition of each of systems ($1,2,...,n-1$) is resonantly coupled to the
cavity mode for an interaction time $t_2=\left( \pi /2\right) /(\sqrt{2J}%
g^{\prime })$ [Fig.~5(c)]. The Hamiltonian describing this step is given by
(in the interaction picture)
\begin{equation}
H_I=g^{\prime }aS^{+}+g^{\prime }a^{+}S^{-}.
\end{equation}
Here and below, $g^{\prime }$ is the resonantly-coupling constant between
the cavity mode and the $\left| 1\right\rangle \leftrightarrow \left|
2\right\rangle $ transition. It can be found that under this Hamiltonian,
the time evolution for the state $\left| J,-J+1\right\rangle \otimes \left|
0\right\rangle _c$ of the systems ($1,2,...,n-1$) and the cavity mode is
described by
\begin{equation}
\left| J,-J+1\right\rangle \otimes \left| 0\right\rangle _c\rightarrow \cos (%
\sqrt{2J}g^{\prime }t)\left| J,-J+1\right\rangle \otimes \left|
0\right\rangle _c-i\sin (\sqrt{2J}g^{\prime }t)\left| J,-J\right\rangle
\otimes \left| 1\right\rangle _c,
\end{equation}
which shows that when systems ($1,2,...,n-1$) are initially in the $W$ state
$\left| J,-J+1\right\rangle ,$ a single photon is created in the cavity mode
after an interaction time $t_2$ given above, through a transformation $%
\left| J,-J+1\right\rangle \otimes \left| 0\right\rangle _c\rightarrow
-i\left| J,-J\right\rangle \otimes \left| 1\right\rangle _c.$ Note that the
operation time $t_2$ ($\propto 1/\sqrt{n-1})$ decreases as the number $n-1$
of the qubit systems increases.

Step (iii): Adjust the level spacings of systems ($1,2,...,n-1$) back to the
original situation [Fig.~5(d)] such that the cavity mode does not couple to
the systems ($1,2,...,n-1$). Meanwhile, adjust the level spacings of system $%
n$ so that the $\left| 1\right\rangle \leftrightarrow \left| 2\right\rangle $
transition of this system is resonantly coupled to the cavity mode for an
interaction time $t_3$ [Fig.~5(d$^{\prime }$)]. The Hamiltonian describing
this step of operation is given by
\begin{equation}
H_I=\hbar \left( g^{\prime }a^{+}\left| 1\right\rangle \left\langle 2\right|
+\text{h.c.}\right) .
\end{equation}
The time evolution of the state $\left| 1\right\rangle _n\left|
1\right\rangle _c$ is described by
\begin{equation}
\left| 1\right\rangle _n\left| 1\right\rangle _c\rightarrow \cos (g^{\prime
}t)\left| 1\right\rangle _n\left| 1\right\rangle _c-i\sin (g^{\prime
}t)\left| 2\right\rangle _n\left| 0\right\rangle _c.
\end{equation}
It can be seen from Eq.~(22) that after an interaction time $t_3=\pi
/(2g^{\prime }),$ the state $\left| 1\right\rangle _n\left| 1\right\rangle
_c $ changes to $-i\left| 2\right\rangle _n\left| 0\right\rangle _c.$ Note
that the state $\left| 0\right\rangle _n\left| 1\right\rangle _c$ remains
unchanged since the state $\left| 0\right\rangle _n$ is not coupled to the
cavity mode. Here and below, the subscript $n$ represents the system $n$.

Step (iv): Leave the level structure of systems ($1,2,...,n-1$) unchanged
[Fig.~5(e)] while adjust the level spacings of system $n$ so that the $%
\left| 0\right\rangle \leftrightarrow \left| 2\right\rangle $ transition of
system $n$ is resonantly coupled to the cavity mode for an interaction time $%
t_4$ [Fig.~5(e$^{\prime }$)]. The Hamiltonian describing this step is
\begin{equation}
H_I=\hbar \left( g^{\prime \prime }a^{+}\left| 0\right\rangle \left\langle
2\right| +\text{h.c.}\right) .
\end{equation}
Here and below, $g^{\prime \prime }$ is the resonantly-coupling constant
between the cavity mode and the $\left| 0\right\rangle \leftrightarrow
\left| 2\right\rangle $ transition. According to this Hamiltonian, one can
easily find that after an interaction time $t_4=\theta /g^{\prime \prime },$
the transformations $\left| 0\right\rangle _n\left| 1\right\rangle
_c\rightarrow \cos \theta \left| 0\right\rangle _n\left| 1\right\rangle
_c-i\sin \theta \left| 2\right\rangle _n\left| 0\right\rangle _c$ and $%
\left| 2\right\rangle _n\left| 0\right\rangle _c\rightarrow -i\sin \theta
\left| 0\right\rangle _n\left| 1\right\rangle _c+\cos \theta \left|
2\right\rangle _n\left| 0\right\rangle _c$ are obtained$.$

The operations for the last three steps are the reverse operations of steps
(i), (ii) and (iii) above, which are described below:

Step (v): Leave the level structure of systems ($1,2,...,n-1$) unchanged
[Fig.~5(d)] while adjust the level spacings of system $n$ such that the $%
\left| 1\right\rangle \leftrightarrow \left| 2\right\rangle $ transition of
system $n$ is resonant with the cavity mode for an interaction time $t_5$
[Fig.~5(d$^{\prime }$)]. The Hamiltonian describing this step is the one in
Eq. (21). The time evolution of the state $\left| 2\right\rangle _n\left|
0\right\rangle _c$ is described by
\begin{equation}
\left| 2\right\rangle _n\left| 0\right\rangle _c\rightarrow \cos (g^{\prime
}t)\left| 2\right\rangle _n\left| 0\right\rangle _c-i\sin (g^{\prime
}t)\left| 1\right\rangle _n\left| 1\right\rangle _c.
\end{equation}
Thus, after an interaction time $t_5=3\pi /(2g^{\prime }),$ the state $%
\left| 2\right\rangle _n\left| 0\right\rangle _c$ becomes $i\left|
1\right\rangle _n\left| 1\right\rangle _c.$ Note that the state $\left|
0\right\rangle _n\left| 1\right\rangle _c$ remains unchanged during this
step of operation.

Step (vi): Adjust the level spacings of system $n$ such that the cavity mode
does not couple to this system [Fig.~5(c$^{\prime }$)]. Meanwhile, adjust
the level spacings of systems ($1,2,...,n-1)$ such that the $\left|
1\right\rangle \leftrightarrow \left| 2\right\rangle $ transition of each of
systems ($1,2,...,n-1$) is resonant with the cavity mode for an interaction
time $t_6$ [Fig.~5(c)]. The Hamiltonian describing this step is the one in
(19), from which one can easily find that the time evolution for the state $%
\left| J,-J\right\rangle \otimes \left| 1\right\rangle _c$ of the systems ($%
1,2,...,n-1$) and the cavity mode is described by
\begin{equation}
\left| J,-J\right\rangle \otimes \left| 1\right\rangle _c\rightarrow \cos (%
\sqrt{2J}g^{\prime }t)\left| J,-J\right\rangle \otimes \left| 1\right\rangle
_c-i\sin (\sqrt{2J}g^{\prime }t)\left| J,-J+1\right\rangle \otimes \left|
0\right\rangle _c.
\end{equation}
It can be seen from Eq.~(25) that the operation of this step results in the
transformation $\left| J,-J\right\rangle \otimes $ $\left| 1\right\rangle
_c\rightarrow $ $-i\left| J,-J+1\right\rangle $ $\otimes \left|
0\right\rangle _c$ for $t_6=\pi /(2\sqrt{2J}g^{\prime }).$

Step (vii): Leave the level structure of system $n$ unchanged [Fig.~5(b$%
^{\prime }$)] while adjusting the level spacings of systems ($1,2,...,n-1$)
such that the $\left| 1\right\rangle \leftrightarrow \left| 2\right\rangle $
transition of each of systems ($1,2,...,n-1$) is non-resonantly coupled to
the cavity mode, with a detuning $\Delta _c$ [Fig.~5(b)]. Then, apply a
pulse to systems ($1,2,...,n-1$) for a duration $t_7=\pi /(2\sqrt{2J}\Omega
) $ [Fig.~5(b)]. As discussed in Sec. III, the operation of this step leads
to the transformation $\left| J,-J+1\right\rangle \rightarrow $ $-i\left|
J,-J\right\rangle .$

Note that after the last step of operation, we will need to leave the level
structure of system $n$ unchanged [Fig.~5(a$^{\prime }$)] while adjust the
level spacings of systems ($1,2,...,n-1$) back to the original situation as
shown in Fig. 5(a), such that the cavity mode does not couple to each system
after the above manipulation.

The states of the whole system after each step of the above operations are
summarized below:
\begin{eqnarray}
&&\ \left| 11...1\right\rangle \left| 0\right\rangle \otimes \left|
0\right\rangle _c  \nonumber \\
&&\ \stackrel{\text{Step(i)}}{\rightarrow }-i\left| J,-J+1\right\rangle
\left| 0\right\rangle \otimes \left| 0\right\rangle _c  \nonumber \\
&&\ \stackrel{\text{Step(ii)}}{\rightarrow }-\left| J,-J\right\rangle \left|
0\right\rangle \otimes \left| 1\right\rangle _c  \nonumber \\
&&\ \stackrel{\text{Step(iii)}}{\rightarrow }-\left| J,-J\right\rangle
\left| 0\right\rangle \otimes \left| 1\right\rangle _c  \nonumber \\
&&\ \stackrel{\text{Step(iv)}}{\rightarrow }-\left| J,-J\right\rangle \left(
\cos \theta \left| 0\right\rangle \left| 1\right\rangle _c-i\sin \theta
\left| 2\right\rangle \left| 0\right\rangle _c\right) {}  \nonumber \\
&&\ \stackrel{\text{Step(v)}}{\rightarrow }-\left| J,-J\right\rangle \left(
\cos \theta \left| 0\right\rangle +\sin \theta \left| 1\right\rangle \right)
\left| 1\right\rangle _c  \nonumber \\
&&\ \stackrel{\text{Step(vi)}}{\rightarrow }i\left| J,-J+1\right\rangle
\left( \cos \theta \left| 0\right\rangle +\sin \theta \left| 1\right\rangle
\right) \left| 0\right\rangle _c  \nonumber \\
&&\ \stackrel{\text{Step(vii)}}{\rightarrow }\left| 11...1\right\rangle
\left( \cos \theta \left| 0\right\rangle +\sin \theta \left| 1\right\rangle
\right) \left| 0\right\rangle _c.
\end{eqnarray}

\begin{eqnarray}
&&\ \left| 11...1\right\rangle \left| 1\right\rangle \otimes \left|
0\right\rangle _c  \nonumber \\
&&\ \stackrel{\text{Step(i)}}{\rightarrow }-i\left| J,-J+1\right\rangle
\left| 1\right\rangle \otimes \left| 0\right\rangle _c  \nonumber \\
&&\ \stackrel{\text{Step(ii)}}{\rightarrow }-\left| J,-J\right\rangle \left|
1\right\rangle \otimes \left| 1\right\rangle _c  \nonumber \\
&&\ \stackrel{\text{Step(iii)}}{\rightarrow }i\left| J,-J\right\rangle
\left| 2\right\rangle \otimes \left| 0\right\rangle _c  \nonumber \\
&&\ \stackrel{\text{Step(iv)}}{\rightarrow }\left| J,-J\right\rangle \left(
\sin \theta \left| 0\right\rangle \left| 1\right\rangle _c+i\cos \theta
\left| 2\right\rangle \left| 0\right\rangle _c{}\right)  \nonumber \\
&&\ \stackrel{\text{Step(v)}}{\rightarrow }-\left| J,-J\right\rangle \left(
-\sin \theta \left| 0\right\rangle +\cos \theta \left| 1\right\rangle
\right) \left| 1\right\rangle _c  \nonumber \\
&&\ \stackrel{\text{Step(vi)}}{\rightarrow }i\left| J,-J+1\right\rangle
\left( -\sin \theta \left| 0\right\rangle +\cos \theta \left| 1\right\rangle
\right) \left| 0\right\rangle _c  \nonumber \\
&&\ \stackrel{\text{Step(vii)}}{\rightarrow }\left| 11...1\right\rangle
\left( -\sin \theta \left| 0\right\rangle +\cos \theta \left| 1\right\rangle
\right) \left| 0\right\rangle _c.
\end{eqnarray}
where $\left| J,-J\right\rangle $ (i.e., $\left| 11...1\right\rangle $) and $%
\left| J,-J+1\right\rangle $ are the Dicke states of systems ($1,2,...,n-1$%
), while $\left| 0\right\rangle ,$ $\left| 1\right\rangle ,$ and $\left|
2\right\rangle $ are the states of system $n.$

On the other hand, it is noted that the following states of the whole system
\begin{equation}
\;\left\{ \left| i_1i_2...i_{n-1}\right\rangle \left| 0\right\rangle
_n\left| 0\right\rangle _c,\;\left| i_1i_2...i_{n-1}\right\rangle \left|
1\right\rangle _n\left| 0\right\rangle _c\right\}
\end{equation}
(for all $i_1,i_2,\cdot \cdot \cdot ,i_{n-1}\in \left\{ 0,1\right\} $ and $%
\prod_{k=1}^{n-1}i_k=0$) remain unchanged during the entire operation. This
is because: (a) During the operation of step (i), the states $\left\{ \left|
i_1i_2...i_{n-1}\right\rangle \right\} $ of systems ($1,2,...,n-1$) were not
affected by the applied pulse, as discussed in Sec. III; and (b) No photon
was emitted to the cavity during the operation of step (ii), when systems ($%
1,2,...,n-1$) are in any one of the states $\left\{ \left|
i_1i_2...i_{n-1}\right\rangle \right\} .$ Hence, it can be concluded from
Eqs.~(26) and (27) that the transformation (2), i.e, the $n$-qubit
controlled-$R$ gate, was implemented with $n$ systems (i.e., the $n-1$
control systems ($1,2,...,n-1$) and the target system $n$) after the above
process.

The systems not involved in each step of the operations above need to be
decoupled from the cavity field and/or the pulse. This requirement can be
achieved by adjusting the level spacings (e.g., doable for superconducting
devices as discussed in the introduction).

The detunings $\Delta _c$ and $\Delta _p$ are set identical for each of
systems ($1,2,...,n-1$) in steps (i) and (vii), and systems ($1,2,...,n-1$)
are brought to resonance with the cavity mode in steps (ii) and (vi).
Therefore, the level spacings for systems ($1,2,...,n-1$) can be
synchronously adjusted via changing the common external parameters of the
qubit systems during the entire operation. In addition, the cavity mode is
virtually excited during the operation of steps (i) and (vii). Thus,
decoherence caused by the cavity decay for these two steps is greatly
reduced.

From the description given above, it can be seen that the level $\left|
0\right\rangle $ of each of qubit systems ($1,2,...,n-1$) is not affected
during the entire operation, because the cavity mode was set to be highly
detuned (decoupled) from the $\left| 0\right\rangle \leftrightarrow \left|
1\right\rangle $ transition and the $\left| 0\right\rangle \leftrightarrow $
$\left| 2\right\rangle $ transition. Thus, the level spacing between the two
levels $\left| 0\right\rangle $ and $\left| 1\right\rangle $ and the level
spacing between the two levels $\left| 0\right\rangle $ and $\left|
2\right\rangle $ are both not required to be identical for each of qubit
systems ($1,2,...,n-1$). However, as shown above, the level spacing between
the two levels $\left| 1\right\rangle $ and $\left| 2\right\rangle $ needs
to be identical for each of qubit systems ($1,2,...,n-1$). Note that for
superconding qubit systems, it is difficult to have the level spacing
between \textit{any} two levels to be the same for each qubit system, but it
is easy to have the level spacing between \textit{certain} two levels (i.e.,
the levels $\left| 1\right\rangle $ and $\left| 2\right\rangle $ for the
present case) to be identical by adjusting device parameters or varying
external parameters [34].

Finally, it should be mentioned that nonuniform of the device parameters for
qubit systems ($1,2,...,n-1$) may cause the coupling strength $g$ or $%
g^{\prime }$ (i.e., the coupling strength between the cavity mode and the $%
\left| 1\right\rangle \leftrightarrow \left| 2\right\rangle $ transition)
not to be the same for each of qubit systems ($1,2,...,n-1$). However, it is
noted that for a superconducting qubit system, the qubit-cavity coupling
strength is adjustable by varying the position of the qubit system in the
cavity. Thus, by having the qubit systems ($1,2,...,n-1$) located at
appropriate positions of the cavity, one can have the coupling strength $g$
or $g^{\prime }$ to be identical for each of qubit systems ($1,2,...,n-1$).

\begin{center}
\textbf{V. POSSIBLE EXPERIMENTAL REALIZATION}
\end{center}

In this section, we give a discussion on possible experimental
implementations. For the method to work:

(a) During the operation of step (i) or step (vii), the occupation
probability $p_1$ of the Dicke state $\left| J,-J+2\right\rangle $ due to
the $\left| J,-J+1\right\rangle \leftrightarrow \left| J,-J+2\right\rangle $
transition induced by the pulse, and the occupation probability $p_2$ of the
Dicke state $\left| J-l/2,-(J-l/2)+1\right\rangle $ due to the $\left|
J-l/2,-(J-l/2)\right\rangle \leftrightarrow \left|
J-l/2,-(J-l/2)+1\right\rangle $ transition induced by the pulse, given by
[35]
\begin{eqnarray}
p_1 &\simeq &\frac{\Omega ^2}{\Omega ^2+\lambda ^2/\left[ 2\left( n-2\right)
\right] },  \nonumber \\
p_2 &\simeq &\frac{\Omega ^2}{\Omega ^2+\left( l\lambda \right) ^2/\left[
4\left( n-l-1\right) \right] }  \nonumber \\
\  &\leq &\frac{\Omega ^2}{\Omega ^2+\lambda ^2/\left[ 4\left( n-2\right)
\right] },
\end{eqnarray}
need to be negligibly small in order to reduce the gate error.

(b) According to the discussion in Sec. III, the following conditions need
to be satisfied:
\begin{equation}
g\ll \Delta _c,\;\,\Omega \sqrt{n-1}\ll g^2/\Delta _c,\;\,\Delta _p=\left(
n-1\right) g^2/\Delta _c.
\end{equation}
Note that these conditions can in principle be achieved because:\ (i) the
Rabi frequency $\Omega $ can be adjusted by changing the intensity of the
pulse, (ii) the detuning $\Delta _c$ can be adjusted by changing the $\left|
1\right\rangle \leftrightarrow \left| 2\right\rangle $ transition frequency $%
\omega _0$, and (iii) the detuning $\Delta _p$ can be adjusted by varying
the pulse frequency $\omega .$

(c) The total operation time is given by
\begin{equation}
\tau =\pi /(\Omega \sqrt{n-1})+\pi /(g^{\prime }\sqrt{n-1})+2\pi /g^{\prime
}+\theta /g^{\prime \prime }+8\tau _a,
\end{equation}
which shows that for a given $\Omega \sqrt{n-1},$ the $\tau $ \textit{%
decreases} as the number $n$ of qubits \textit{increases}. Here, $\tau _a$
is the typical time required for adjusting the level spacings during each
step. The $\tau $ should be much shorter than the energy relaxation time $%
\gamma _{2r}^{-1}$ and dephasing time $\gamma _{2p}^{-1}$ of the level $%
\left| 2\right\rangle ,$ such that decoherence, caused due to spontaneous
decay and dephasing process of the qubit systems, is negligible during the
operation. And, the $\tau $ needs to be much shorter than the lifetime of
the cavity photon, which is given by $\kappa ^{-1}=Q/2\pi \nu _c,$ such that
the decay of the cavity photon can be neglected during the operation. Here, $%
Q$ is the (loaded) quality factor of the cavity and $\nu _c$ is the cavity
field frequency. To obtain these requirements, one can design the qubit
systems to have sufficiently long energy relaxation time and dephasing time,
such that $\tau \ll $ $\gamma _{2r}^{-1},\gamma _{2p}^{-1};$ and choose a
high-$Q$ cavity such that $\tau \ll \kappa ^{-1}.$

\begin{figure}[tbp]
\includegraphics[bb=46 272 550 682, width=8.6 cm, clip]{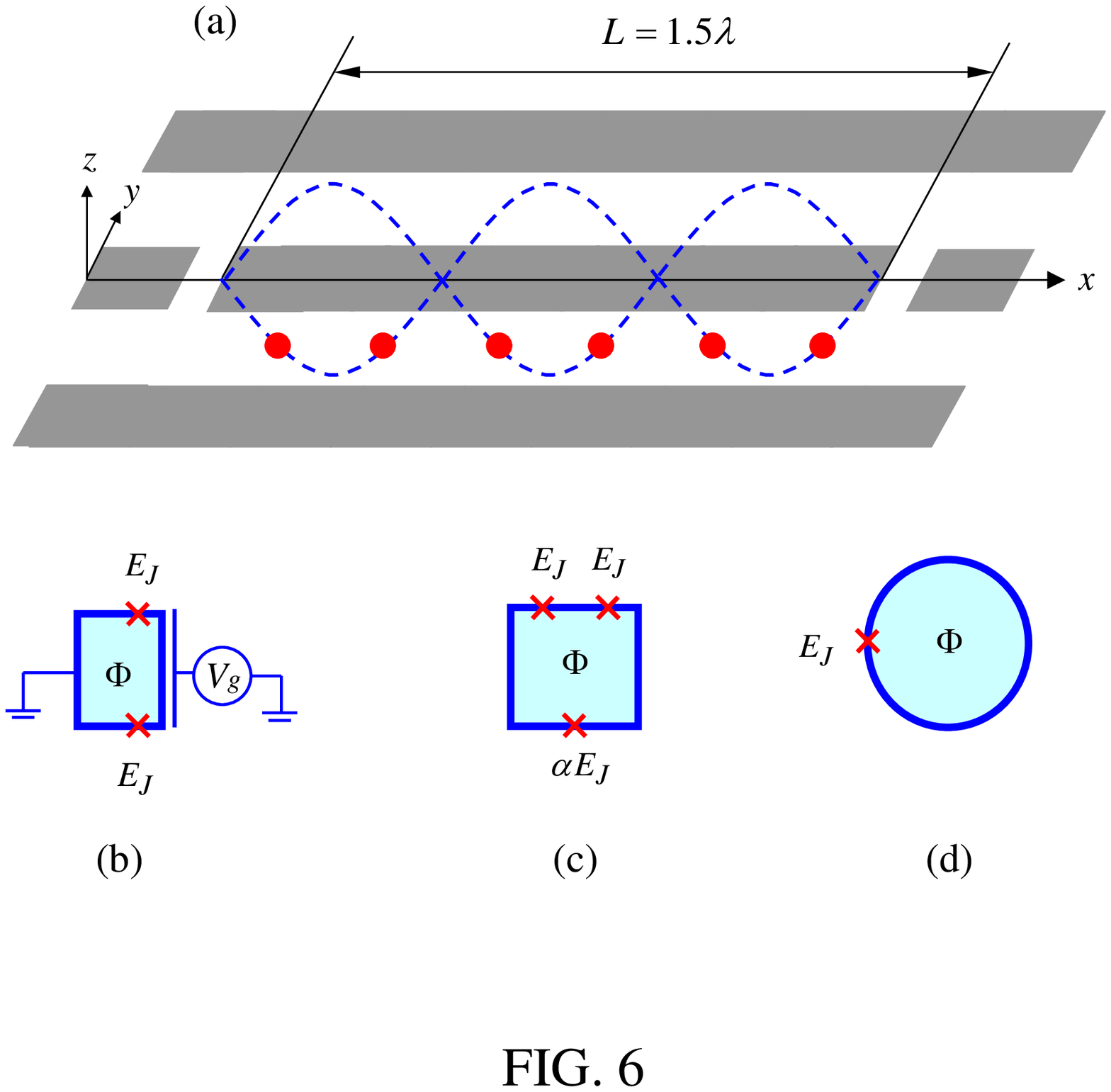} %
\vspace*{-0.08in}
\caption{(Color online) (a) Sketch of the setup for six superconducting
qubit systems (red dots) and a (grey) standing-wave quasi-one-dimensional
coplanar waveguide resonator. $\lambda $ is the wavelength of the resonator
mode, and $L$ is the length of the resonator. The two blue curved lines
represent the standing wave magnetic field, which is in the $z$-direction.
Each qubit system (a red dot) could be a superconducting charge-qubit system
shown in (b), flux-qubit system in (c), and flux-biased phase-qubit system
in (d). The qubit systems are placed at locations where the magnetic fields
are the same to obtain an identical coupling constant for each qubit system.
The superconducting loop of each qubit system, which is a large square for
(b) and (c) while a large circle for (d), is located in the plane of the
resonator between the two lateral ground planes (i.e., the $x$-$y$ plane). $%
E_J$ is the Josephson junction energy ($0.6<\alpha <0.8$) and $V_g$ is the
gate voltage. In addition, the external magnetic flux $\Phi $ applied to the
superconducting loop for each qubit system is created by the magnetic field
threading the superconducting loop.}
\label{fig:6}
\end{figure}

For the sake of definitiveness, let us consider the experimental possibility
of realizing a six-qubit controlled-Hadamard gate (i.e., the controlled-$R$
gate for $\theta =\pi /4$), using six identical superconducting qubit
systems coupled to a resonator [Fig.~6(a)]. Each qubit system could be a
superconducting charge-qubit system [Fig.~6(b)], flux-qubit system
[Fig.~6(c)], or flux-biased phase-qubit system [Fig.~6(d)]. As a rough
estimate, assume $g/{2\pi }\sim 220$ MHz, which could be reached for a
superconducting qubit system coupled to a one-dimensional standing-wave CPW
(coplanar waveguide) transmission resonator [36]. With the choice of $%
g^{\prime },g^{\prime \prime }\sim g$, $\Delta _c\sim 10g,$ $\Omega /2\pi
\sim 1.1$ MHz (i.e., $\lambda /\Omega \sim 20$)$,$ and $\tau _a\sim 1$ ns,
one has $\tau \sim 0.2$ $\mu $s, much shorter than $\min \{\gamma
_{2r}^{-1},\gamma _{2p}^{-1}\}\sim 1$ $\mu $s [26,37]. In addition, consider
a resonator with frequency $\nu _c\sim 3$ GHz (e.g., Ref. [38]) and $Q\sim
5\times 10^4$, we have $\kappa ^{-1}\sim 2.7$ $\mu $s, which is much longer
than the operation time $\tau $ here. Note that superconducting coplanar
waveguide resonators with a quality factor $Q>10^6$ have been experimentally
demonstrated [39].

For the choice of $\Delta _c\sim 10g$ and $\lambda /\Omega \sim 20$ here, we
have $p_1\sim 0.02$ and $p_2\leq 0.04,$ which can be further reduced by
increasing the ratio $\Delta _c/g$ and $\lambda /\Omega .$ How well this
gate would work needs to be further investigated for each particular
experimental set-up or implementation. However, we note that this requires a
rather lengthy and complex analysis, which is beyond the scope of this
theoretical work.

\begin{center}
\textbf{VI. CONCLUSION}
\end{center}

In summary, we have proposed a way for implementing an $n$-qubit
controlled-rotation gate with three-level superconducting qubit systems in
cavity QED. This proposal requires seven steps of operation only, which is
independent of the number $n$ of qubits. In contrast, when the proposed gate
is constructed by using the conventional gate-decomposing protocols, the
basic operational steps increase with the number $n$ of qubits \textit{%
exponentially}. Thus, when the number $n$ of qubits is large, the gate
operation is significantly simplified by using the present proposal. In
addition, as shown above, the gate operation time for this proposal
decreases as the number of qubits increases. This proposal is quite general,
which can be applied to various types of superconducting devices in a cavity
or coupled to a resonator.

\begin{center}
\textbf{ACKNOWLEDGMENTS}
\end{center}

C.P.Y. is grateful to Shi-Biao Zheng for very useful comments. This work is
supported in part by the National Natural Science Foundation of China under
Grant No. 11074062, the Zhejiang Natural Science Foundation under Grant No.
Y6100098, funds from Hangzhou Normal University, and the Open Fund from the
SKLPS of ECNU.

\end{document}